\documentclass[10 pt,conference]{IEEEtran}
\usepackage{amssymb,amsmath}
\usepackage{graphicx}
\usepackage[export]{adjustbox}
\graphicspath{ {images/} }

\newlength{\qedlengte}
\newcommand{\qed}{\hspace*{1em}\hfill\qedbox}

\def\blackbox{{\rule{2.00mm}{2.00mm}}}
\def\qed{\hspace*{\fill}\blackbox}

\begin{document}

\title{Three-Way Joins on MapReduce:\\ An Experimental Study}


\author{
\IEEEauthorblockN{Ben Kimmett}
\IEEEauthorblockA{University of Victoria, BC, Canada \\
blk@uvic.ca}
\and
\IEEEauthorblockN{Alex Thomo}
\IEEEauthorblockA{University of Victoria, BC, Canada \\
thomo@cs.uvic.ca}
\and
\IEEEauthorblockN{S. Venkatesh}
\IEEEauthorblockA{University of Victoria, BC, Canada \\
venkat@cs.uvic.ca}
}

\date{}

\maketitle

\begin{abstract}
We study three-way joins on MapReduce. 
Joins are very useful in a multitude of applications 
from data integration and traversing social networks, 
to mining graphs and automata-based constructions.  
However, joins are expensive, even for moderate data sets; we need efficient algorithms to perform distributed computation of joins using clusters of many machines. 
MapReduce has become an increasingly popular distributed computing system and programming paradigm.  
We consider a state-of-the-art MapReduce multi-way join algorithm by Afrati and Ullman and show when it is appropriate for use on very large data sets. By providing a detailed experimental study, we demonstrate that this algorithm scales much better than what is suggested by the original paper. 
However, if the join result needs to be summarized or aggregated, as opposed to being only enumerated, then the aggregation step can be integrated into a cascade of two-way joins, making it more efficient than the other algorithm, and thus becomes the preferred solution.
\end{abstract}

\section{Introduction}
\label{intro}

The importance of joins cannot be overstated. 
Joins are used explicitly or implicitly when performing a multitude of everyday tasks, such as connecting two or more datasets based on common attributes, comparing tuples with other tuples in the same table using selfjoins, traversing graphs (c.f.~\cite{Gephi,MST07,GT06,KSKRA11}), mining graphs and social networks (c.f.~\cite{WWZSPYH05,CF06,XYFS07,KT13}) computing graph statistics and cubes (c.f.~\cite{THP08,ZTP10,ZLXH11}), and
multiplying matrices (c.f.~\cite{ML12}), to name a few.
Joins are so useful, they find application even in algorithms requiring the computation of intersection of large automata and transducers (c.f.~\cite{TVY08,ST11}), or probabilistic reasoning on graph databases (c.f.~\cite{HST13}).
In fact, it is hard to imagine a domain where joins are {\em not} present, albeit sometimes in disguise. This is because of our innate need, in many fields of research, to always be able to connect different pieces of data or information together.

In many of the aforementioned settings, the datasets involved are very big.  For example, Facebook, the most popular social network, contains data for over 900 million users and their relationships. To traverse the Facebook graph of friendships and compute statistics would involve a series of challenging multi-way joins.
The main reason for the difficulty is that joins are expensive, even for datasets of moderate size. As such, devising distributed algorithms for computing joins on large data sets is of the utmost importance.

MapReduce (c.f.~\cite{DG04,DG08,DG10}) is a popular distributed computing framework that can work with thousands of machines in a fault-tolerant way. Unsurprisingly, joins have been one of the first candidates to be considered for implementation in MapReduce. While joining two tables of data is easy to implement from an algorithmic point of view, joining three or more tables brings several challenges to overcome. 

In this paper, we focus on  three-way joins. 
Afrati and Ullman in \cite{AU10,AU11} give an elegant algorithm for computing three-way and multi-way joins in MapReduce. However, as we describe later, their main idea for computing the join is based on the assumption of having a limited number of processes (reducers) for processing intermediate results. This is a limiting factor for the scalability of a multi-way join for large clusters with thousands of machines. 
Also, more often than not, one is more interested in summarizing or aggregating the result of the join in some way. The algorithm proposed in \cite{AU10,AU11} needs to produce the entire join result before an aggregator can summarize it. In contrast, a simple cascade of two way joins may be a better choice, as it allows interleaving the aggregation with the computation of the intermediate result. 

We outline the details of how to perform this optimization, and then perform a detailed experimental study on the performance of the algorithm of \cite{AU10,AU11} versus a simple cascade of two-way joins. Our results reveal two surprising facts: 
\begin{enumerate}
\item For real data (such as those coming from edge-lists of social networks), the algorithm of \cite{AU10,AU11} can scale on clusters much bigger than the original papers suggest, and 
\item When aggregation is required (rather than enumerating the raw join result) a cascade of two-way joins is the preferred choice exhibiting a significant gain in execution cost.  
\end{enumerate}

We pay special attention to applying joins for multiplication of large, sparse matrices because of the wide applicability 
in social network and web analysis. This is also in line with the prime use cases of \cite{AU10,AU11}, 
which also come from the analysis of large social graphs. 
Such graphs are often represented as sparse matrices
by listing the non-empty elements of their incidence matrix (the so-called edge list table).


\medskip
The outline of the paper is as follows: 
In Section~\ref{jmg}, we formally describe joins, and then matrix multiplication and graph computations based on joins.
In Section~\ref{mapreduce} we describe the MapReduce framework. 
In Section~\ref{threewayjoins} we discuss three-way join algorithms for MapReduce. 
In Section~\ref{aggregation} we give aggregation algorithms for join results. 
In Section~\ref{eval} we present our experimental evaluation.
Section~\ref{conclusions} concludes the paper.

\section{Joins, Matrices, Graphs}
\label{jmg}

Let $R(A,B,V)$ and $S(B,C,W)$, be two tables with attributes (columns) $A$, $B$, $V$, and $B$, $C$, $W$, respectively. 
The join $R(A,B,V) \Join S(B,C,W)$ is table 
\begin{eqnarray*}
J(A,B,C,V,W) &=& \{(a,b,c,v,w): \\ 
										& & \;\;\;\; (a,b,x) \in R(A,B,V) \mbox{ and } \\
										 & & \;\;\;\; (b,c,y) \in S(B,C,W)\}.
\end{eqnarray*}
The definition is extended in the natural way when $A$, $B$, $C$, $V$, and $W$ are sets of attributes, 
as opposed to single attributes.

The above is also called a ``two-way'' join because {\em two} tables are joined.
If three tables are joined, we refer to the join as being ``three-way''.
Join is an associative operation, i.e. 
$(R(A,B,V) \Join S(B,C,W))\Join T(C,D,X)$ $=$ 
$R(A,B,V) \Join (S(B,C,W))\Join T(C,D,X))$.

With minimal work, the concept of a join can be extended to perform matrix multiplication. 
A table $R(A,B,V)$ represents a sparse matrix, 
with each tuple $(a,b,v)$ in $R$ representing the presence of $v$ in the matrix, at row $a$ and column $b$. 

When two matrix tables $R(A,B,V)$ and $S(B,C,W)$ are ``joined''
the effect is to perform the first step of matrix multiplication. 
$R$ and $S$ are joined on their common attribute, $B$, 
and the $V$ and $W$ values are multiplied. Call the result $J(A,C,P)$. 
Observe that in the result we only keep $A$ and $C$, and multiply the values of $V$ and $W$ rather than 
just output them as in the join definition.
In effect, the row vectors of the matrix represented by $R$ have been multiplied by the column vectors of the matrix represented by~$S$.

The next step is to perform summation of the $\mbox{$p=v\cdot w$}$ values of $J(A,C,P)$ tuples that agree on $A$ and $C$, 
and produce one tuple $(a,c,s_{a,c})$ for each existing $a,c$ combination in $J(A,C,P)$.
That is, we group by $a,c$ and aggregate using sum. 
This is the equivalent of summing the intermediate results of matrix multiplication to yield the finished matrix. 

Now, graphs can be represented as incidence matrices, which are often quite sparse for real graphs. 
We can use join-based matrix multiplication to multiply graph matrices with themselves $n$ times  
and thus obtain the number (or the weight) of paths of length $n+1$ 
between the starting and ending nodes listed in the final output.
This is important in the friend-of-friend analysis of social networks.

Also, by considering the diagonal of the result (those $(a,c,s_{a,c})$ tuples with $a=c$) 
for binary incidence matrices,
we can obtain the number of triangles in the graph. 
Namely, the number of triangles is the sum of all $s_{a,c}$, with $a=c$, 
divided by three.

As the matrix multiplication is a simple extension of the join followed by a group by and aggregation, 
we will first focus on MapReduce algorithms for join, then modify them to handle matrix multiplication.

\section{MapReduce}
\label{mapreduce}

MapReduce is used to describe both a distributed computing system and an algorithmic paradigm, 
used to process large sets of data.
The most popular incarnation of MapReduce as a distributed system is Hadoop; an open-source Java implementation based on the Hadoop Distributed File System (HDFS). 

From an algorithmic point of view, MapReduce simplifies distributed computing. 
All a programmer needs to do is implement 
a Map and a Reduce function, without having to deal with low level details 
of machine communication, data transfer, scheduling, and fault-tolerance. 
Depending on the number of machines and configuration of the system, 
the Map and Reduce functions can run in many machines at once. 

There are always two phases in a MapReduce job, the Map phase and the Reduce phase. The latter starts once the former completes.
The processes running the map function are called {\em mappers}, and
the processes running the reduce function are called {\em reducers}.

Both the map and reduce functions take {\em key-value pairs} (KVPs) as input. They also emit key-value pairs; the exact construction of any KVP depends on the individual map or reduce function used.
The records emitted by the mappers are sorted and shuffled by the system before being sent to the reducers.
The guarantee of the system is that all the emitted KVPs with the same key are sent to the same reducer. A reducer can receive KVPs with many different keys, but if it receives one KVPs with a specific key, it is certain to receive all other KVPs with that same key. 

A two-way join $R(A,B,V) \Join S(B,C,W)$ can be implemented in MapReduce as follows.
Initially, each mapper is assigned a chunk of data, which can contain tuples from $R(A,B,V)$, $S(B,C,W)$, or both. 
The specification for the Map and Reduce functions is as follows.

\noindent
{\bf Map function.} \\
For each pair $(tid, (a,b,v))$, where $(a,b,v) \in R(A,B,V)$, emit $(b, (a,v,R))$.\\ 
For each pair $(tid, (b,c,w))$, where $(b,c,w) \in S(B,C,W)$, emit $(b, (c,w,S))$.

\noindent
{\bf Reduce function.}\\
Join all $(a,b,v)$'s from $(b, (a,v,R))$ \\
with all $(b,c,w)$'s from $(b, (c,w,S))$, matching on $b$. \\
Emit $((a,b,c,v,w),\dag)$, where $\dag$'s value is unimportant. 
Some attributes may optionally be omitted from the output.
  
\medskip
Because of the system guarantee that all KVPs with the same key are sent to the same reducer, 
the reducer receiving the $(b, (\_,\_,\_))$ pairs has all the information it needs to compute the section of the join 
related to value $b$.

What really matters for the efficiency of a MapReduce algorithms is the total amount of I/O performed by all the processes. Emitted data is considered I/O; this is because data emission and transmission is realized using HDFS. The total amount of I/O is called the {\em communication cost}.
However, as typical in databases when we compare algorithms, we do not count the cost of writing the final output. 
The reasoning behind this is that the output is either small enough to be 
consumed by human users (in which case the cost can be safely ignored), or if not, it will be pipelined into another process (possibly another MapReduce round) which will summarize or aggregate it in some way. In this case, the size of the output would be counted in the communication cost of the next process, and so is not included in this cost estimate.

The communication cost for the above join is $2r+2s$, 
where $r$ and $s$ are the sizes of $R$ and $S$, respectively. 
This is because each tuple of $R$ and $S$ will be read once by a mapper, and the results (which are the same size, as the mapper emits a single tuple for each one it reads) will be read again, with each modified tuple of $R$ and $S$ read by a single reducer.

\section{Three-Way Joins}
\label{threewayjoins}

Computing the three way join $R(A,B,V) \Join S(B,C,W) \Join T(C,D,X)$ is harder, and can be done several different ways.
The simplest way way is to do a cascade of two two-way joins: first, compute $R(A,B,V) \Join S(B,C,W)$, then join the result with $T(C,D,X)$. 
Since this algorithm uses two rounds of MapReduce, we call it the {\em two-round three-way join} algorithm, or (2,3J).
The communication cost for the cascade cannot be determined beforehand because 
the size of the intermediate join $R(A,B) \Join S(B,C)$ cannot be known before we compute it, 
but once this cost is known, the cost of the cascade is $2r+2s+2t+2\lvert R\Join S\rvert$, 
where $r$, $s$, and $t$ are the sizes of $R$, $S$, and $T$. 

An alternate algorithm for three-way joins is proposed by 
Afrati and Ullman in~\cite{AU10,AU11}. 
This algorithm uses only one round of MapReduce; therefore, we call it the {\em one-round three way join}, or 1,3J. 
In the 1,3J algorithm, the number of reducers to be used is an explicit parameter $k=k_1k_2$
(in fact two parameters $k_1$ and $k_2$ whose product is $k$). 
We create two hash functions, $h$ and $g$, that hash to $k_1$ and $k_2$ buckets, respectively. 
Then the map and reduce functions are as follows. 

\noindent
{\bf Map function.} 
\begin{itemize}
\item For each $(tid, (b,c,w))$ from $S$, \\
emit $((h(b),g(c)),(S,(b,c,w)))$. 
\item For each $(tid, (a,b,v))$ from $R$, \\
emit $((h(b),j),(R,(a,b,v)))$, for $j\in [1,k_2]$.
\item For each $(tid, (c,d,x))$ from $T$, \\
emit $((i,g(c)),(T,(c,d,x)))$, for $i\in [1,k_1]$.
\end{itemize}

\noindent
{\bf Reduce function.} \\
Join all $(b,c,w)$'s from $((h(b),h(c)),(S,(b,c,w)))$'s \\
with all $(a,b,v)$'s from $((h(b),h(c)),(R,(a,b,v)))$'s \\
and all $(c,d,x)$'s from $((h(b),g(c)),(T,(c,d,x)))$'s, \\
matching on $b$ and $c$, respectively, and emitting $((a,b,c,d,v,w,x),\dag)$. 
Some attributes may optionally be omitted.

The mappers will emit one key-value pair for each tuple of the middle table $S$, 
and $k_1$ and $k_2$ KVPs for each tuple of $T$ and $R$, respectively. 
Let $(b,c,w)$ be a tuple in $S$.
The reducer receiving KVPs with the key $((h(b),h(c))$ will have all the information to construct the part of the three-way join due to $(b,c,w)$. 

\vspace{5 pt}
\begin{figure}
\centering
\includegraphics[scale=0.7]{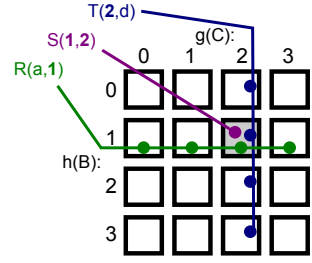}
\caption{This chart demonstrates how the 1,3J algorithm passes data to reducers. 
KVPs from $R$ and $T$ are sent to a 'row' and 'column' of reducers, respectively, 
while tuples from $S$ go to a single reducer to join them together. 
}
\end{figure}

The communication cost of this algorithm is
$(r+s+t) + (s + k_1t + k_2r)$, 
where $r+s+t$ is the cost  
to read the chunks of $R$, $S$, and $T$ once, 
$s$ is the cost to move each $(b,c)\in S$ to the reducer handling $(h(b),h(c))$, 
$k_1t$ is the cost to move each $(c,d)\in T$ to the reducers handling $(i,g(c))$, for $i\in [1,k_1]$, and
$k_2r$ is the cost to move each $(a,b)\in R$ to the reducers handling $(h(b),j)$, for $j\in [1,k_1]$.

As shown in~\cite{AU10,AU11}, this cost is minimized when $k_1=\sqrt{kr/t}$ and $k_2=\sqrt{kt/r}$. For these values, the communication cost is
$r+2s+t+2\sqrt{krt}$. When we join a table to itself (a selfjoin), such as when computing the third power of a graph adjacency matrix, we have $r=s=t$, so $k_1=k_2=\sqrt{k}$, and the communication cost equals $4r+2r\sqrt{k}$.  

\medskip
\noindent
{\bf Remark 1.} 
It is not always clear when, and if, 1,3J is better than 2,3J. 
While the 2,3J algorithm can generate a significant cost when it produces intermediate results, the 1,3J algorithm generates multiple KVPs for each tuple in $R$ and $T$. The more reducers are used in the computation, the more KVPs 1,3J will generate; however, 2,3J will always generate the same number of KVPs given the same input.
Thus, there is a scalability problem with 1,3J. After a certain point, 
more machines do not necessary mean better efficiency. 
For a realistic example considered in \cite{AU10,AU11}, 
1,3J's communication cost surpasses 2,3J after reaching 960 reducers. 
Using today's multicore commodity machines of 8 cores, this translates into only 120 machines.
Nonetheless, we show experimentally in the next section that for datasets derived from real-world graphs, 
the critical number of reducers is typically far greater. 
Therefore, 1,3J wins over 2,3J for modest clusters of computers when the goal is to only to {\em enumerate} the result. 
However, this may not be the goal of several common applications.  

\medskip
\noindent
{\bf Remark 2.}
More often then not, we do not consume the result of the join directly, 
but instead summarize or aggregate it in some way, e.g. summation in matrix multiplication or counting in statistical applications. 
When employing 1,3J, we need to wait for the join to be fully computed, 
and only then may we apply the aggregation. In contrast, if we use 2,3J, we can run the aggregation in stages, applying it to the intermediate two-way join as described in the next section. As we show experimentally, such an optimization yields significant gains in terms of communication cost.

\section{Aggregation}
\label{aggregation}

Here we will use join-based matrix multiplication as an example. Aggregations for other problems are similar. 
2,3J with aggregation, call it 2,3JA, requires the use of another MapReduce round to serve as the aggregator. 

The aggregator we used after computing the first two-way join $R(A,B,V) \Join S(B,C,W)$ is as follows:

\noindent
{\bf Map function.} \\ 
For all tuples $((a,b,c,v,w), \dag)$, emit $((a,c),p)$, where $p=v\cdot w$.

\noindent
{\bf Reduce function.} \\
For all tuples $((a,c),p)$, sum all the tuples' $p$ values, returning $((a,c),s_{a,c})$.

The same aggregator is also used after the second join. 


The aggregator employed after computing the three-way join using 1,3J is as follows.

\noindent
{\bf Map function.} \\ 
For all tuples $((a,b,c,d,v,w,x), \dag)$, emit $((a,d),p)$, where $p=v\cdot w\cdot x$.

\noindent
{\bf Reduce function.} \\
For all tuples $((a,d),p)$, sum all the tuples' $p$ values, returning $((a,d),s_{a,d})$.

We call 1,3J followed by this aggregator 1,3JA.

%
%
%

2,3JA computes $R(A,B,V) \Join S(B,C,W)$ [of size $r'$], then aggregates the result, 
yielding $Agg(R(A,B,V) \Join S(B,C,W))$ [of size $r''$]. 
This is then joined with $T(C,D,X)$. 
The communication cost is similar to that of 2,3J, but with the addition of $2r''$, 
for a total of $6r+2r'+2r''$ tuples. 
The exact $r''$ depends on how well the aggregator can reduce $R(A,B,V) \Join S(B,C,W)$, 
but it is always equal to or less than (usually much less than) $r'$.

Unlike 2,3J and 2,3JA,
the communication cost of 1,3J for 1,3JA rises with the number of reducers. 
Recall, 1,3J's communication cost is $4r+2r\sqrt{k}$ tuples, where $k$ is the number of reducers.
The computation cost of 1,3JA is $4r+2r\sqrt{k}+2r'''$, where $r'''$ is the size of the raw three-way join.

\section{Evaluation}
\label{eval}

Our experiments ran on a 33-node (4 cores per node, maximum 132 MapReduce instances running at any one time) cluster, using Apache Hadoop 1.2.1. For data, we acquired seven datasets from the Stanford Large Network Dataset Collection (http://snap.stanford.edu/data/). Each dataset represented a directed graph.
Of the datasets, five (Slashdot, Twitter, Wikitalk, Pokec, and LiveJournal) were social in nature, representing user relationships (Twitter, Pokec, Livejournal) or interactions (voting on other users for Slashdot, comments on talk pages for Wikitalk). The remaining two datasets each represented different data: the Amazon set was derived from Amazon's "customers who bought item A also bought item B" database, while the Google Web dataset is a small chunk of internet structure (with edges representing links between pages), which was released by Google as part of a programming competition. 

In all experiments, each dataset was joined to itself twice, creating a three-way selfjoin. 
The three copies of the dataset will still be referred to as $R$, $S$, and $T$.





\subsection{Results}
\label{results}

\begin{figure}
\begin{tabular}{cc}
  \includegraphics[width=4cm]{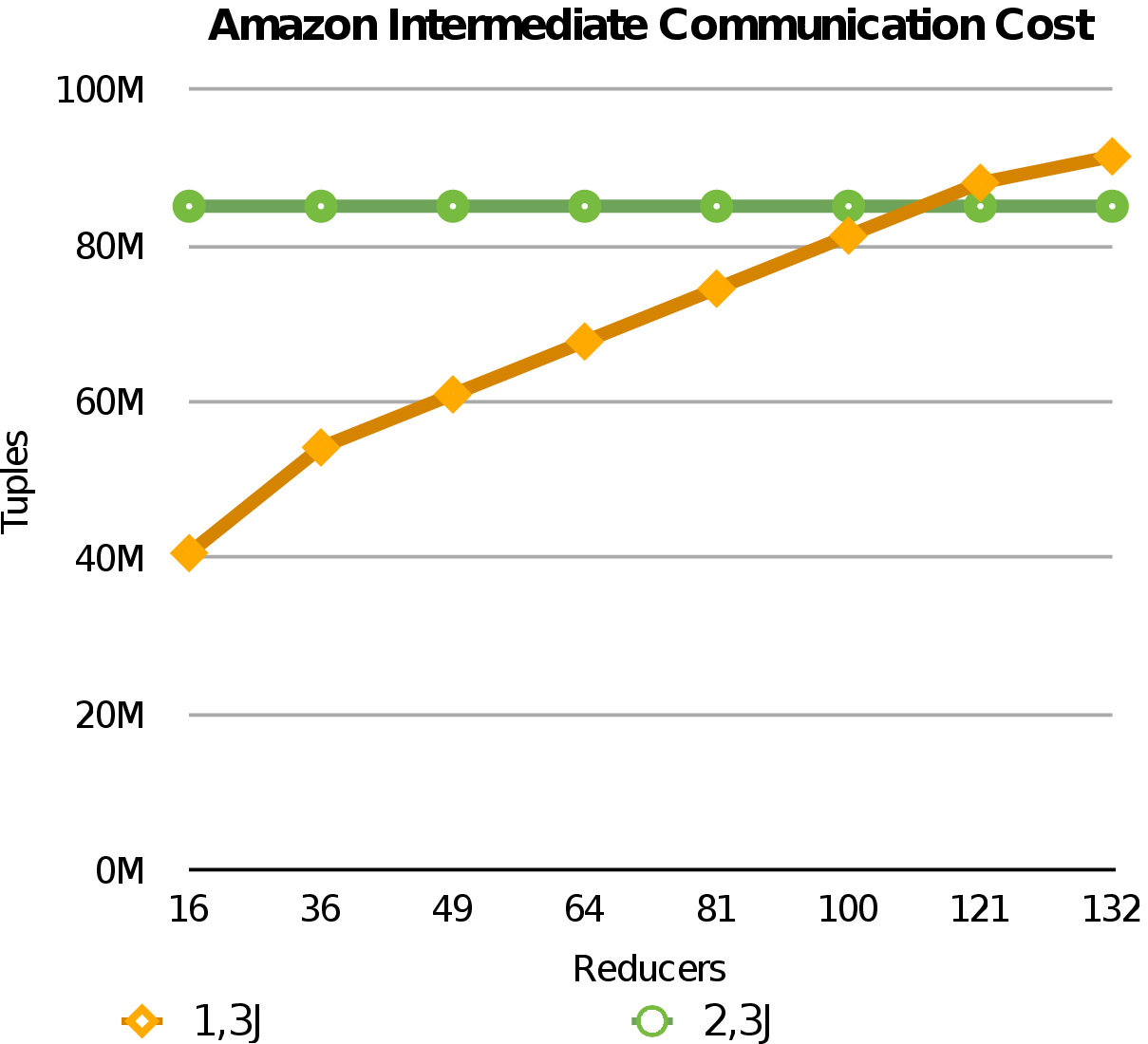} &
  \includegraphics[width=4cm]{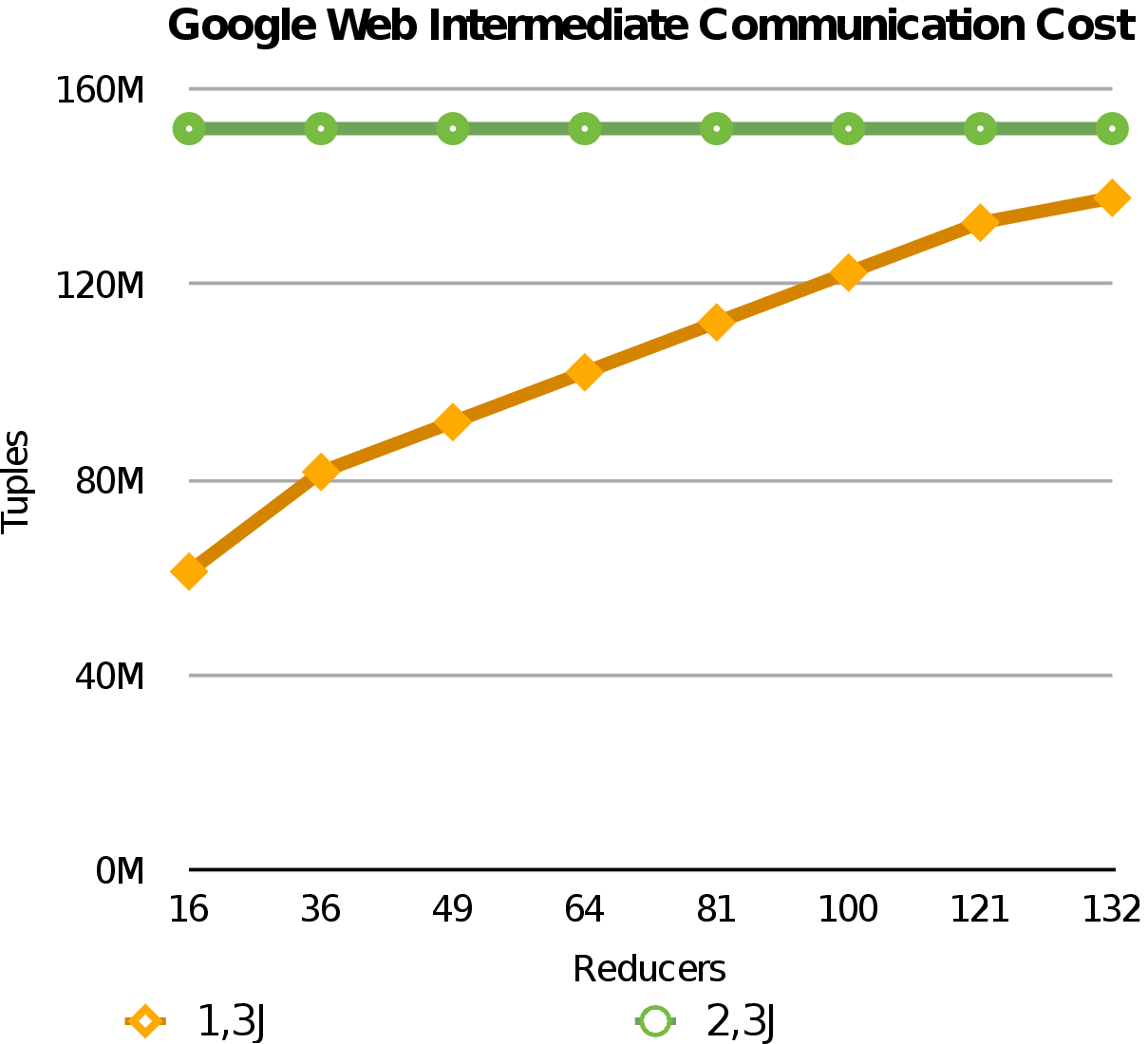} \\\\
  \includegraphics[width=4cm]{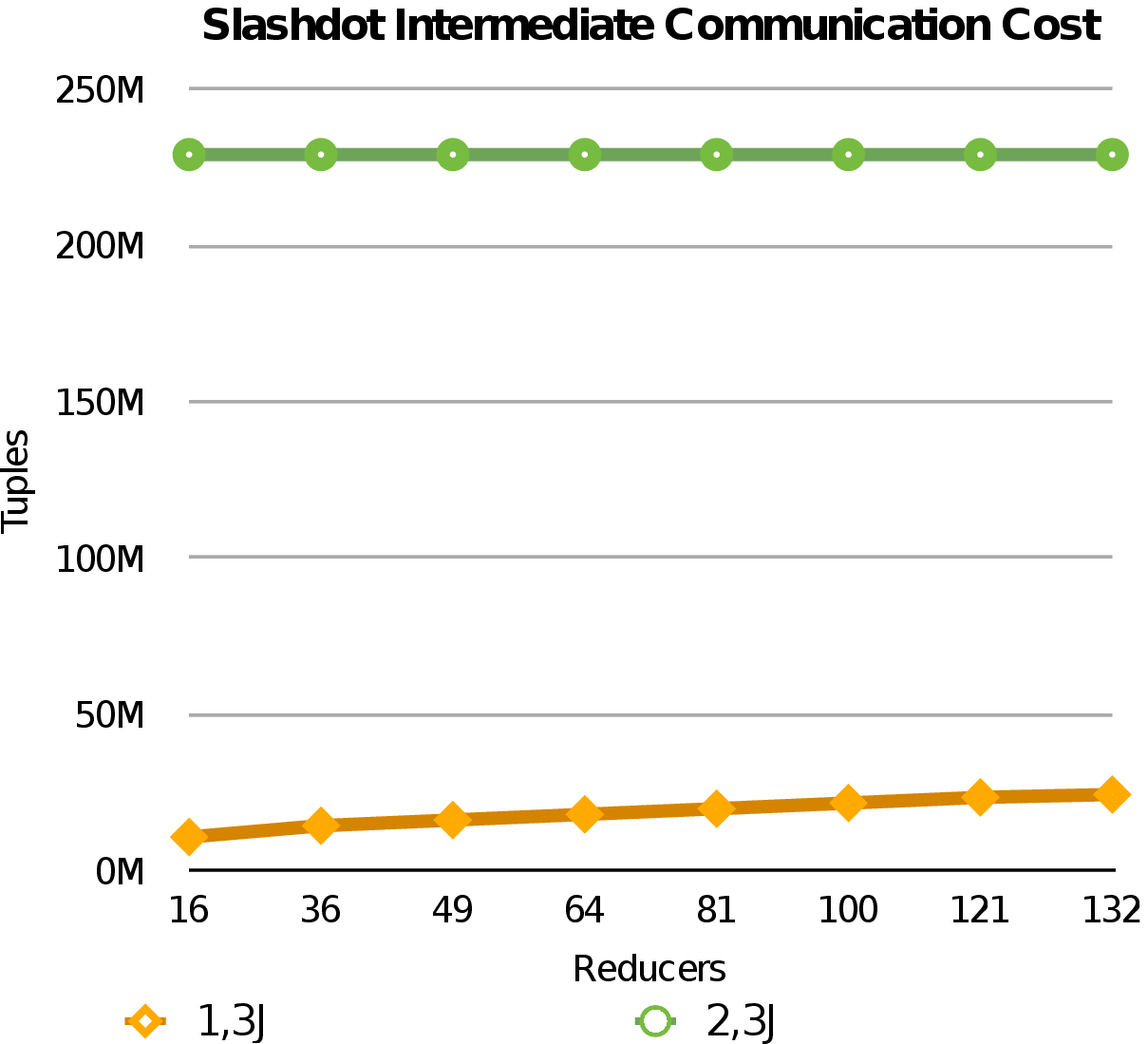} &
  \includegraphics[width=4cm]{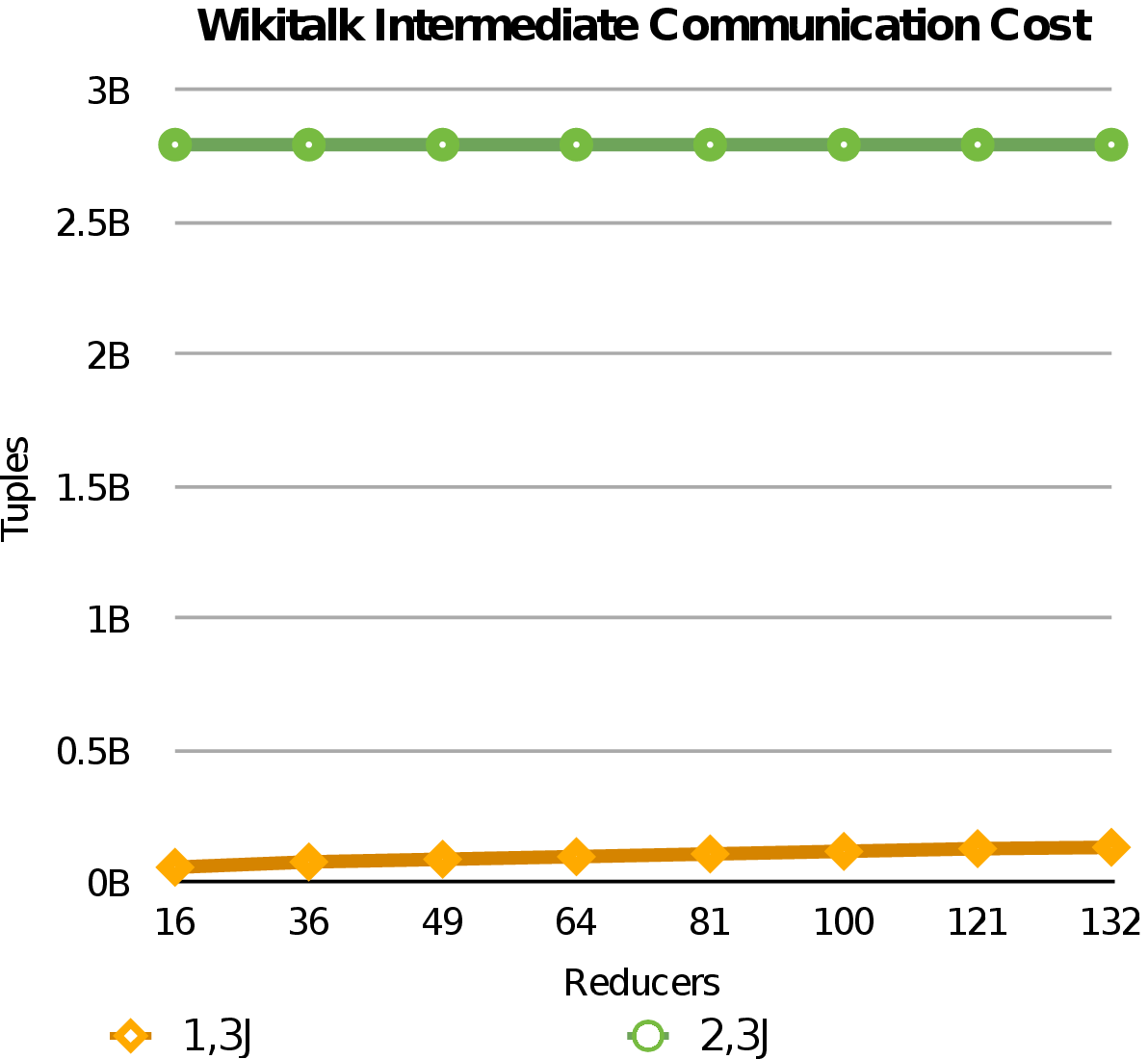} \\\\
   \includegraphics[width=4cm]{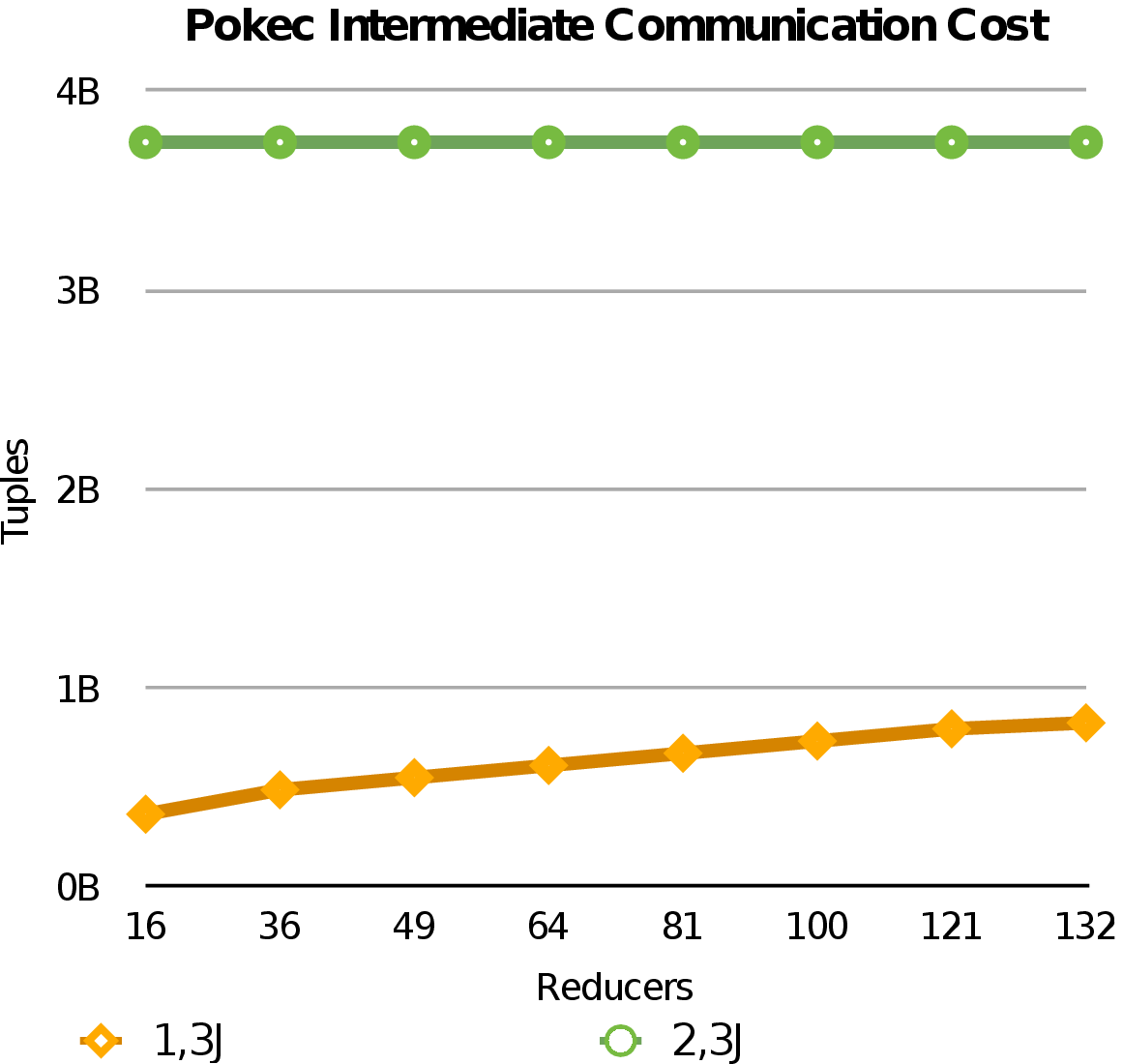} &
  \includegraphics[width=4.1cm]{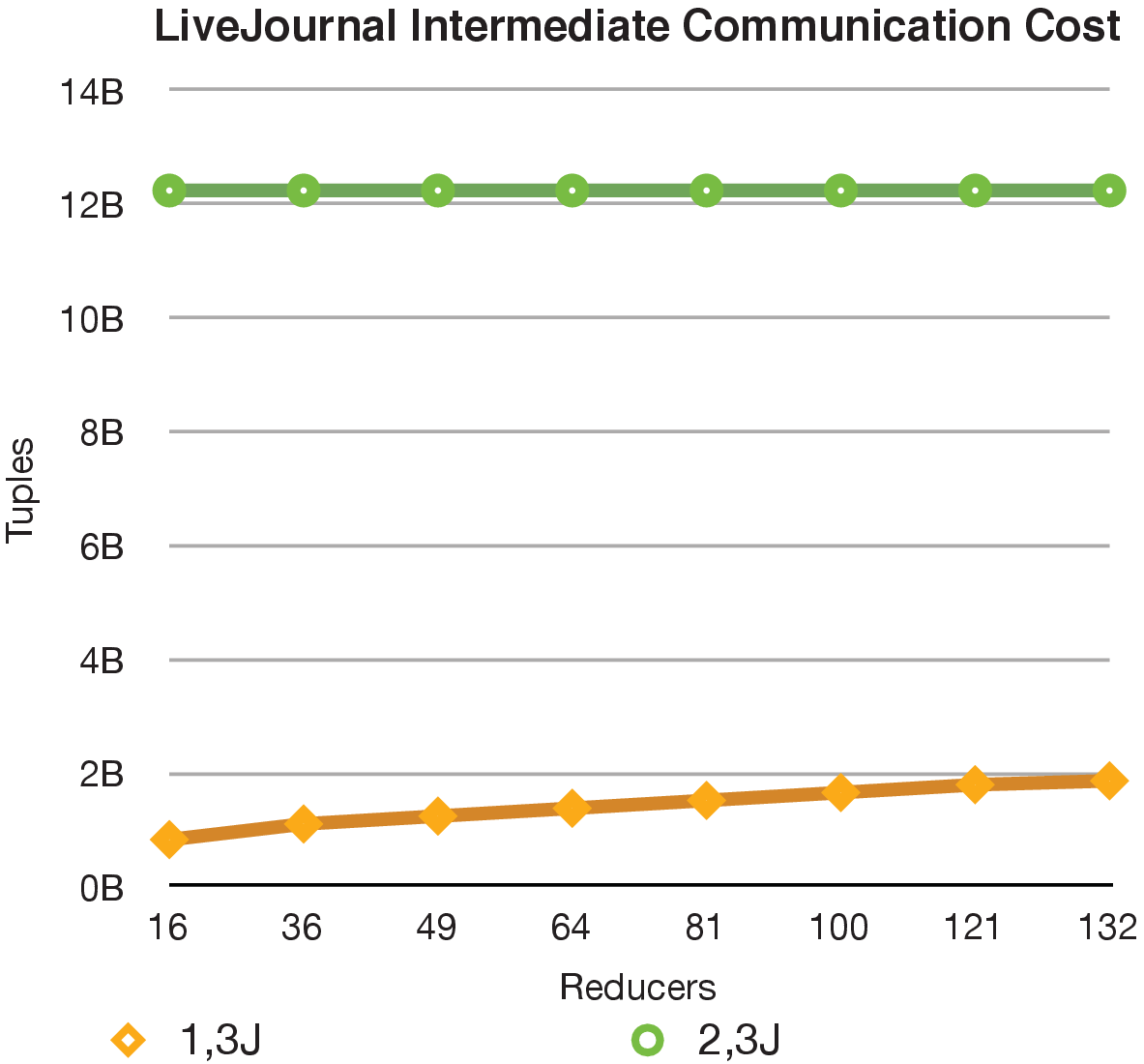}\\\\ 
   \includegraphics[width=4cm]{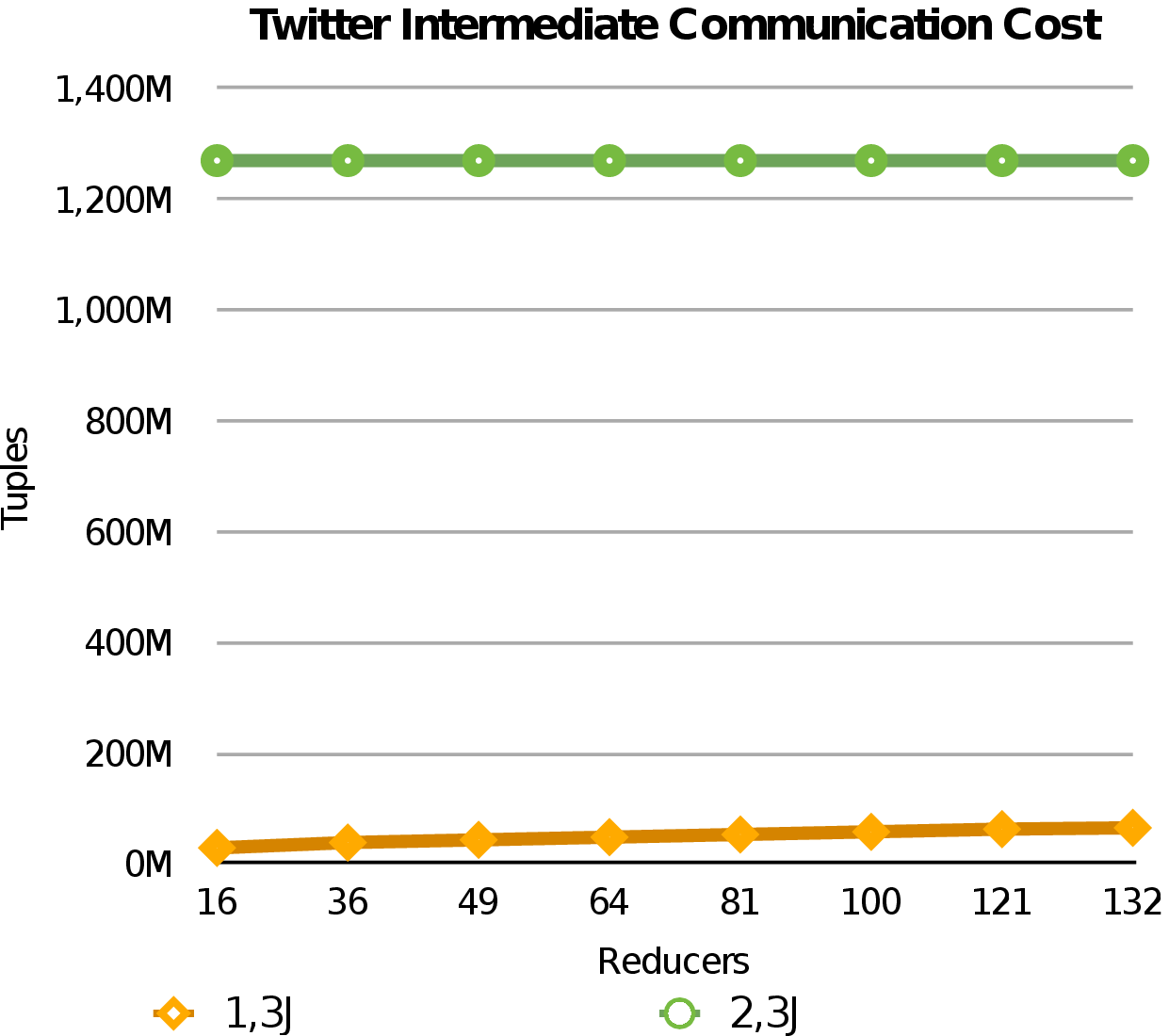} 
\end{tabular}
  \caption{Sizes of intermediate 1,3J and 2,3J communication cost, compared. Top row: Amazon (left), Google Web (right). 2nd row: Slashdot (left), Wikitalk (right). 3rd row: Pokec (left), LiveJournal (right). Bottom left: Twitter.}
  \label{fig:intermed}
\end{figure}

For the first set of tests, we ran our implementation of 2,3J and 1,3J on each dataset. 
We measured the communication cost as defined above. 
The results are shown in Figure~\ref{fig:intermed}.

\noindent

\begin{figure}
\includegraphics[width=8.5cm]{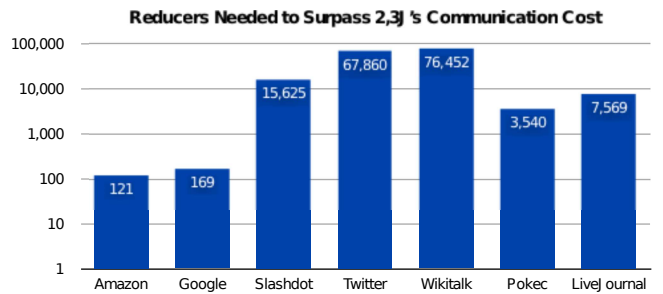}
  \caption{For the 1,3J's intermediate communication cost to surpass the 2,3J's on a specific dataset, it would have to use the listed number of reducers.}
  \label{fig:barchart}
\end{figure}

For every dataset, 
1,3J had a lower communication cost than 2,3J for a large number of reducers. After a certain reducer threshold, the 2,3J cost became lower than the 1,3J cost, but the number of reducers the 1,3J would need to use to cost more than the 2,3J is typically very large, as shown in Figure~\ref{fig:barchart}.
This is a surprising fact that shows that 1,3J can scale much more that what \cite{AU10,AU11} suggest 
(only 960 reducers for a large hypothetical social network). 

As shown in Figure~\ref{fig:intermed}, the Twitter dataset's communication cost was far lower when running the 1,3J algorithm. A cluster running the two algorithms would need to have 67,860 reducers 
(a 260x261 reducer array, or about 8,400 8-core machines) before the 1,3J's communication cost would be larger than 
the 2,3J's. 
Similarly, the LiveJournal dataset would have to be run on a cluster with 7,569 reducers 
(an 87x87 reducer array, about 950 8-core machines) before the cost of running the 1,3J algorithm upon it 
would grow greater than the cost of running the 2,3J on the same. 



\begin{figure}
\begin{tabular}{cc}
  \includegraphics[width=4cm]{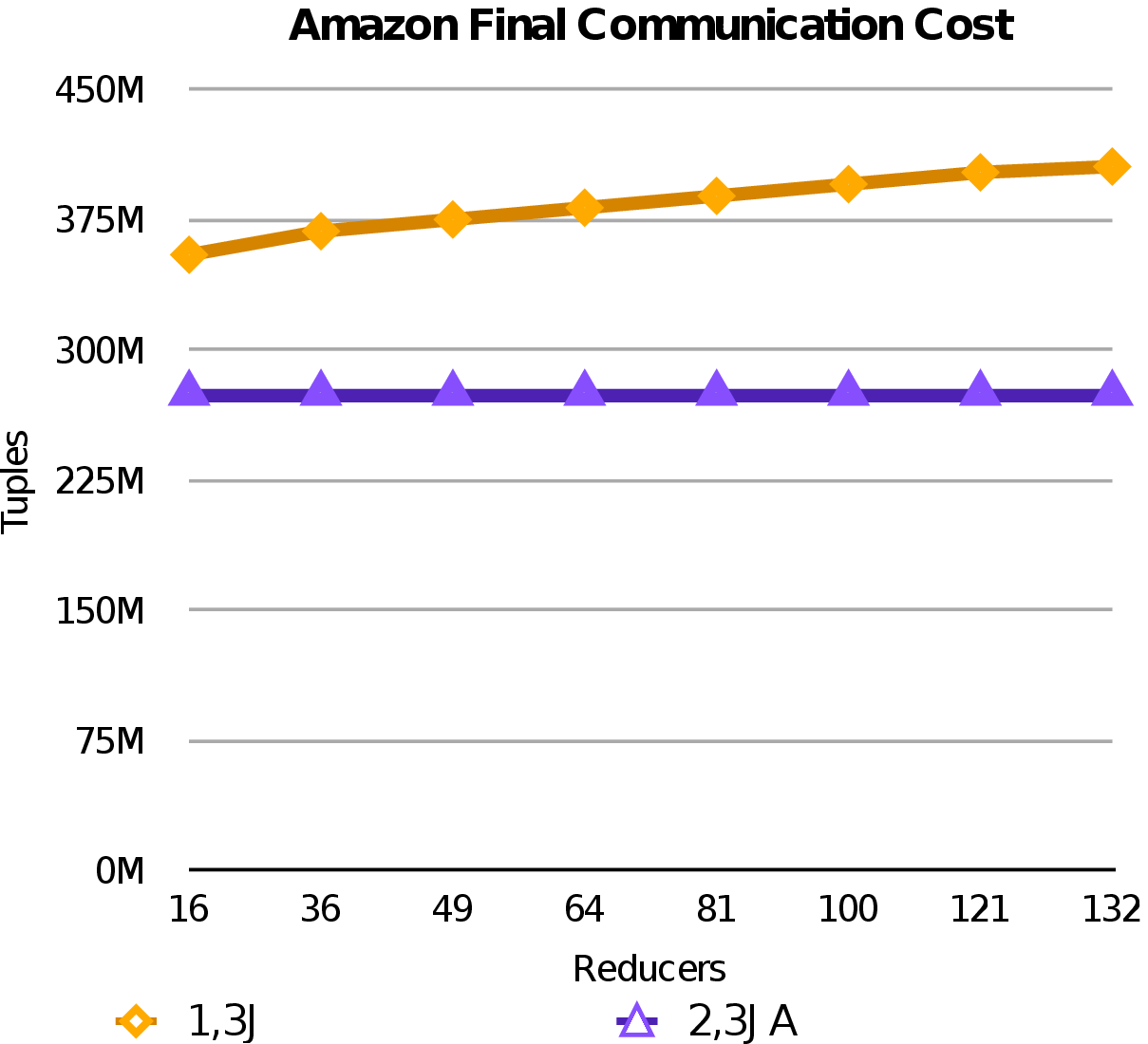} &
  \includegraphics[width=4cm]{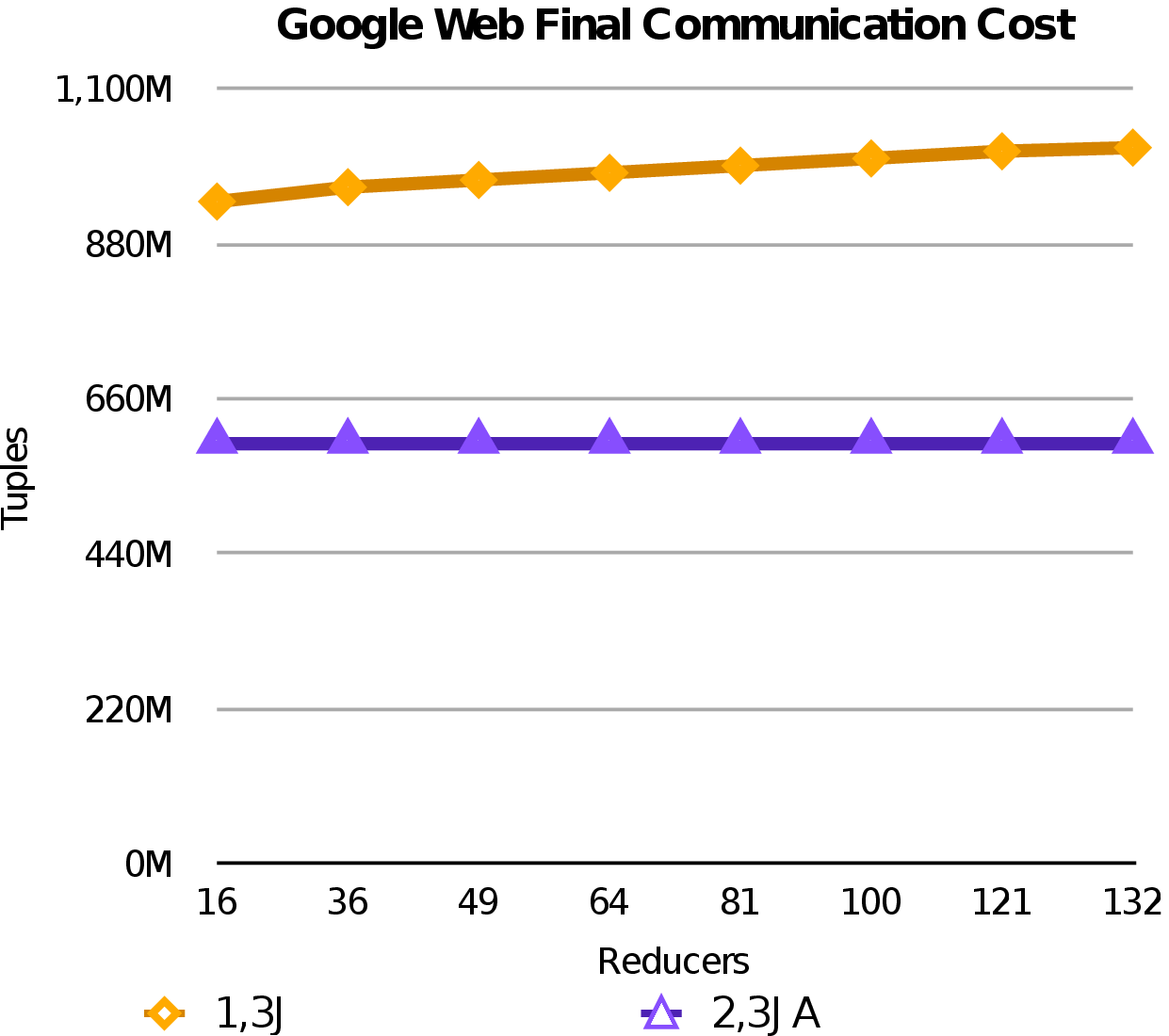} \\\\
  \includegraphics[width=4cm]{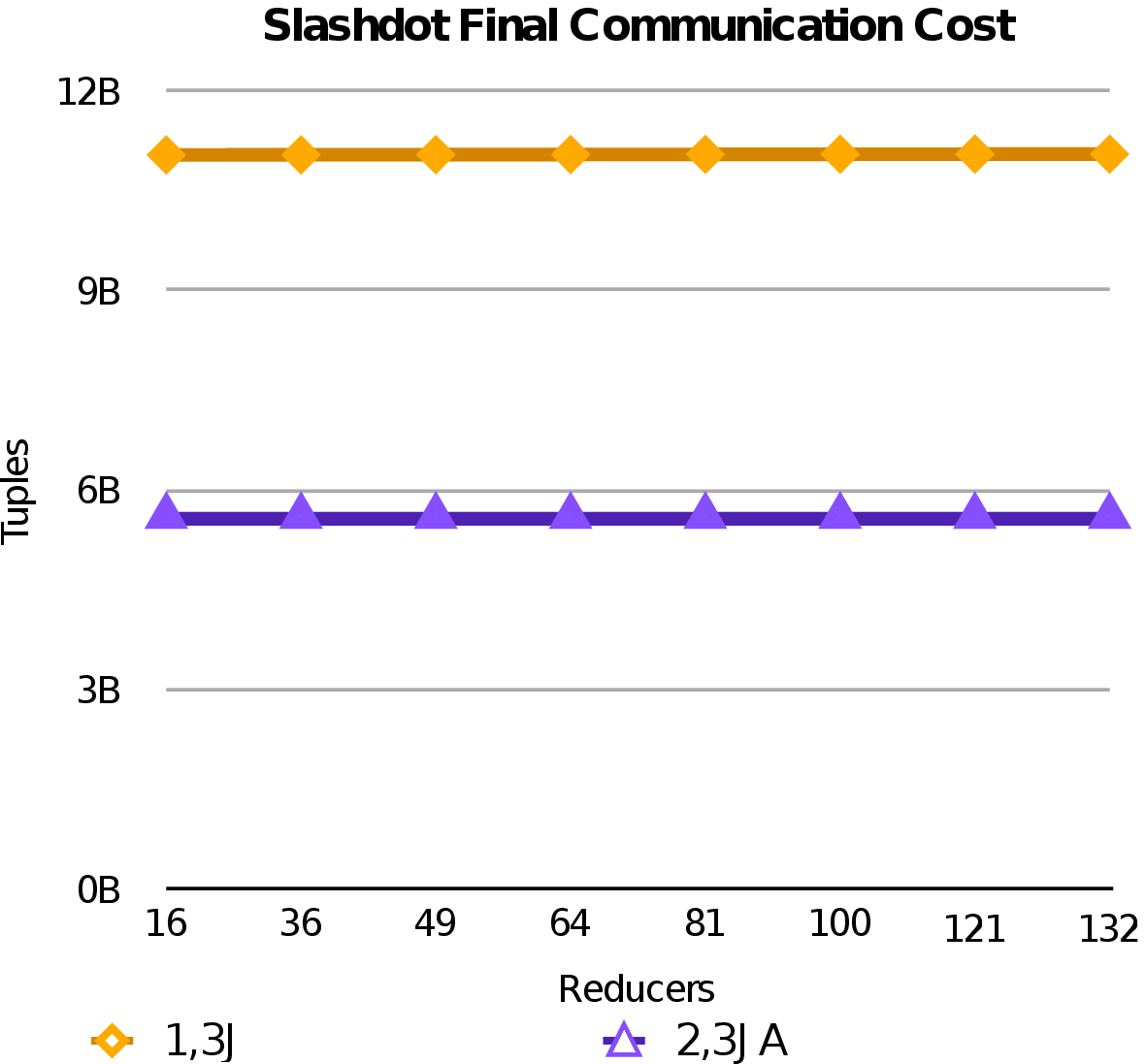} &
   \includegraphics[width=4cm]{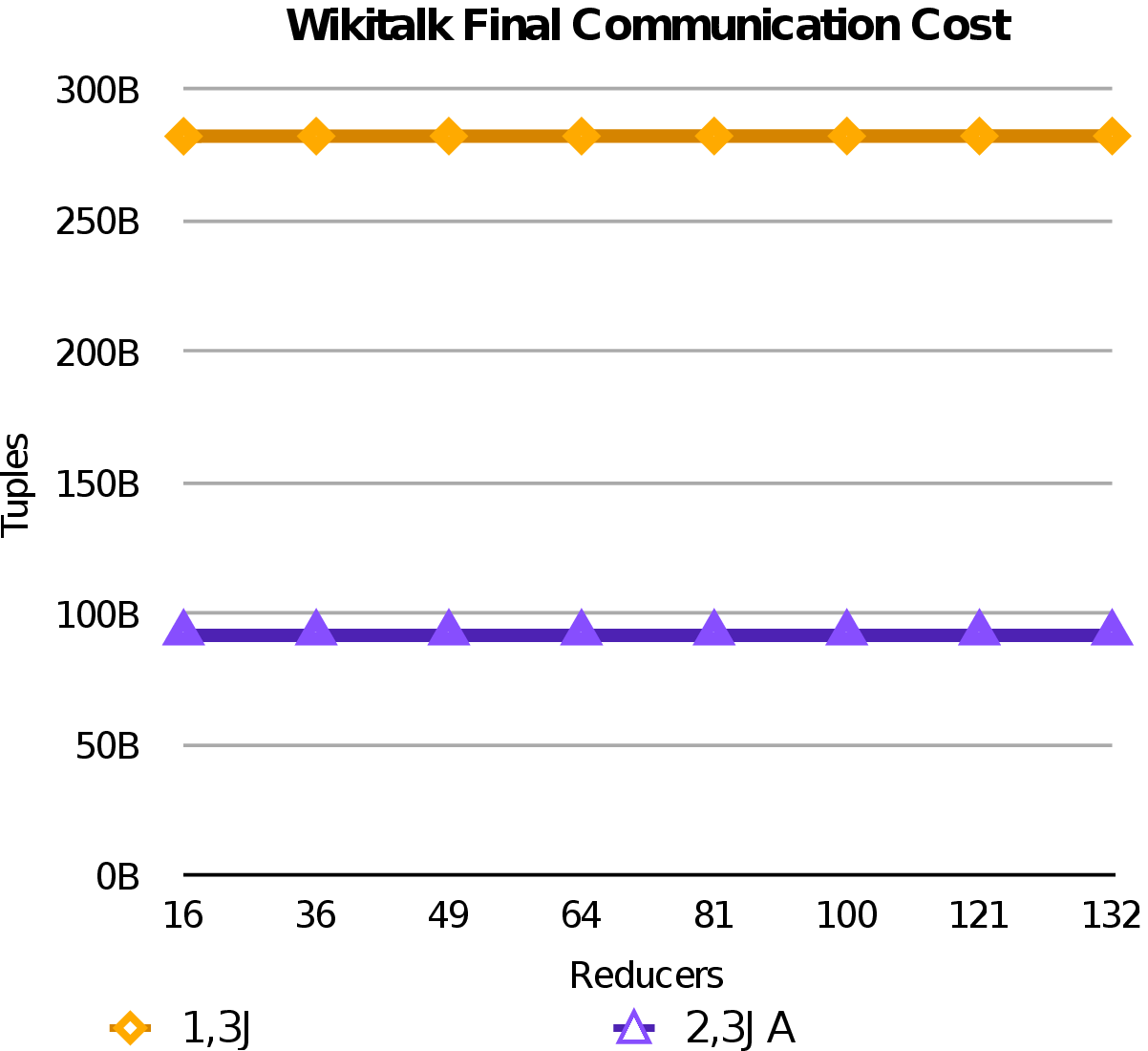} \\\\
  \includegraphics[width=4cm]{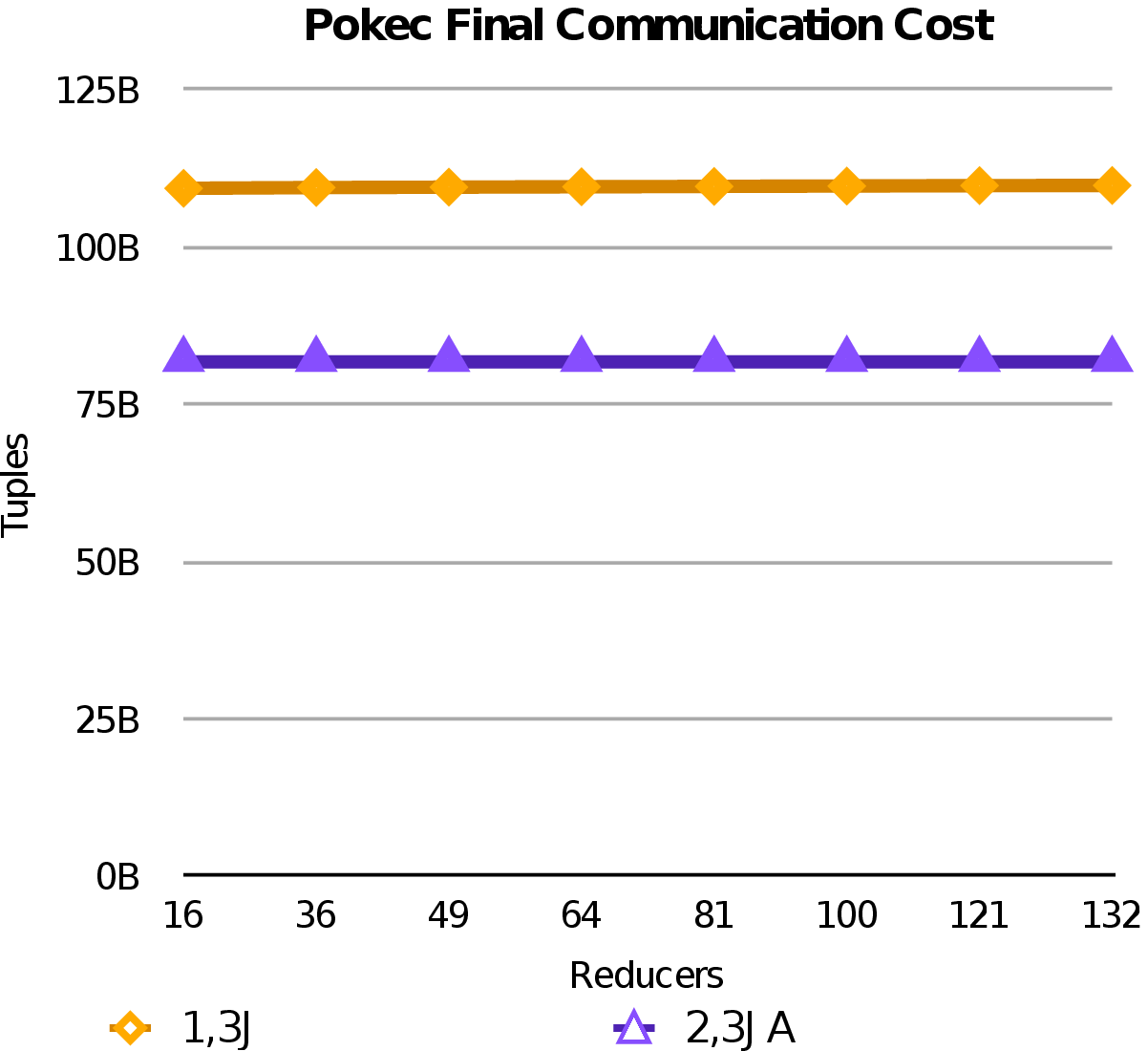} &
   \includegraphics[width=4cm]{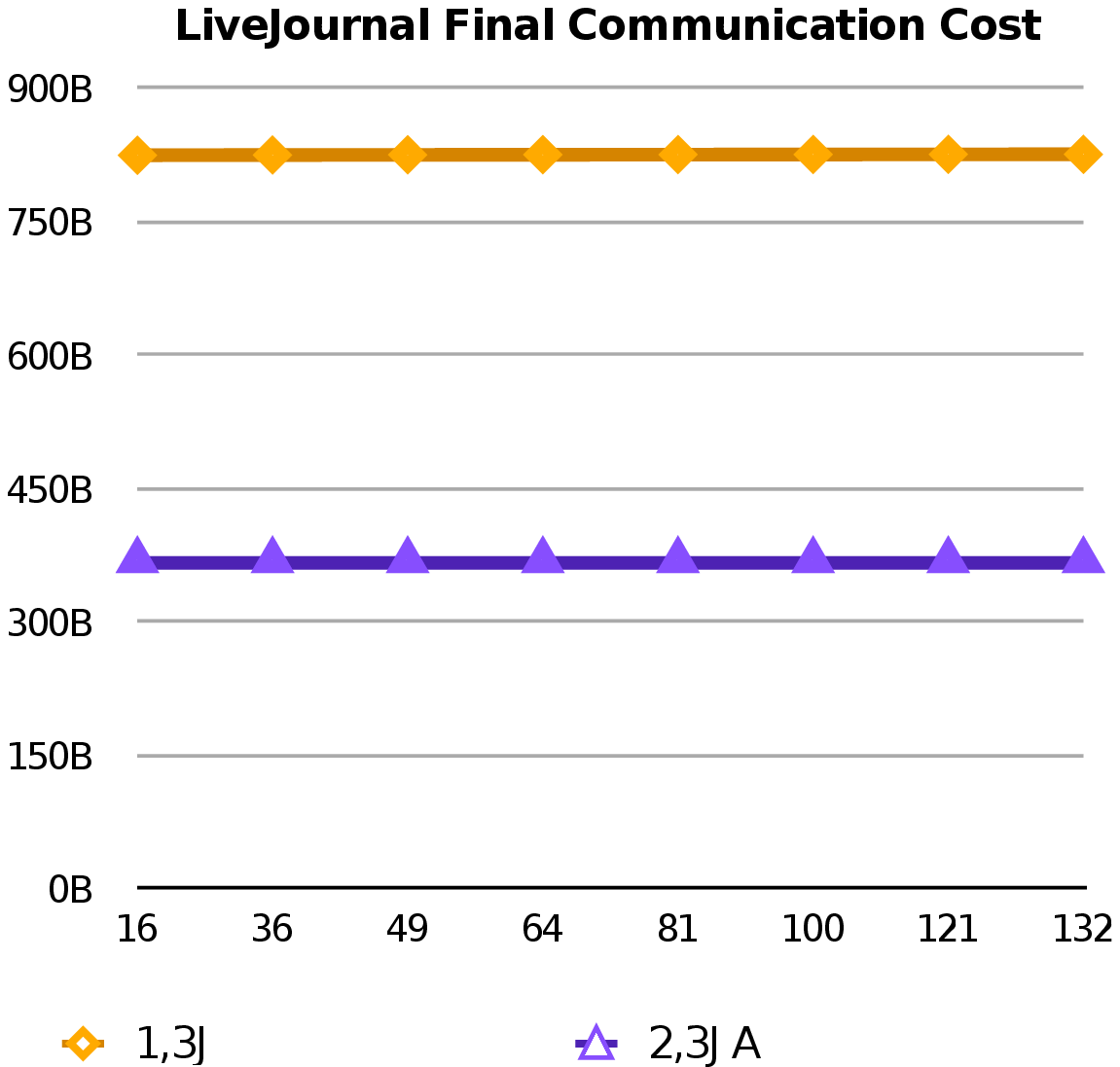} \\\\
    \includegraphics[width=4cm]{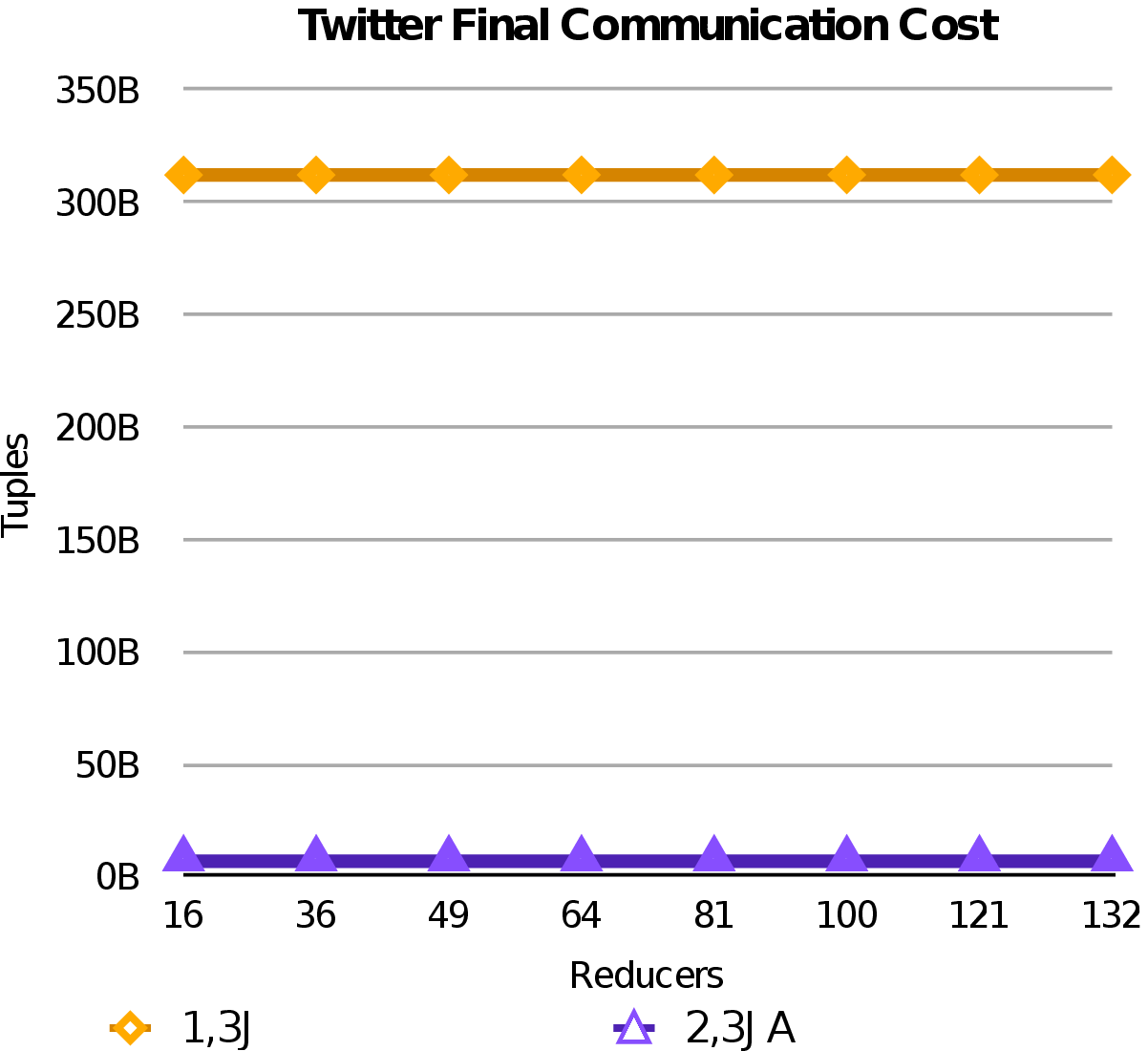} &
   \includegraphics[width=4cm]{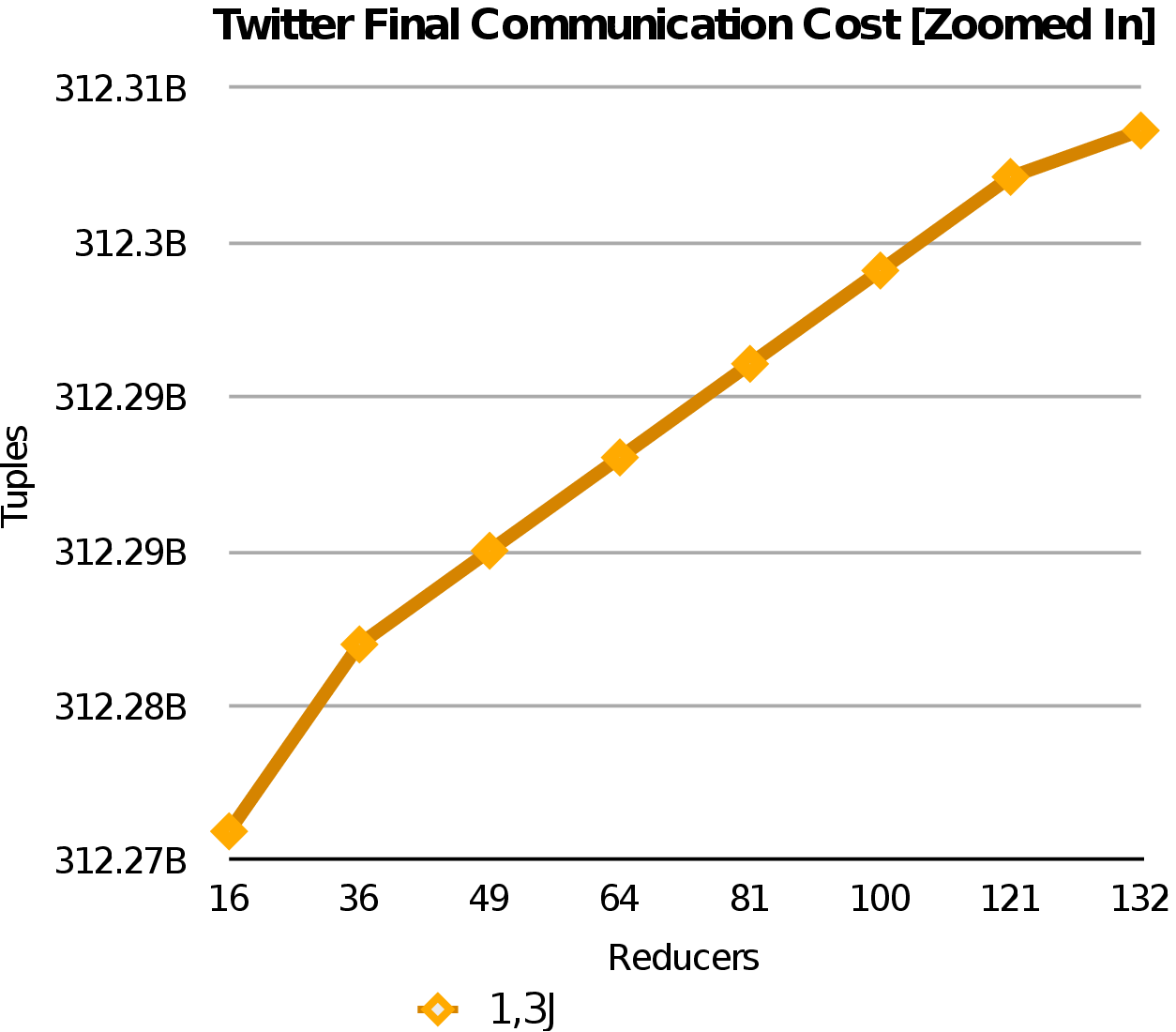}
\end{tabular}
  \caption{Sizes of communication cost of 1,3JA and 2,3JA. Top row: Amazon (left), Google Web (right). 2nd row: Slashdot (left), Wikitalk (right). 3rd row: Pokec (left), LiveJournal (right). Bottom: Twitter (left), Twitter graph scaled to show slope of 1,3J line (right).}
  \label{fig:final}
\end{figure}

For the second set of tests, we compared 2,3JA and 1,3JA on each dataset, 
and the results are shown in Figure~\ref{fig:final}. 
On the graphs of the larger datasets, the 1,3JA line still has a slope (the 2,3JA line is flat), 
but it cannot be easily observed due to the graph's scale; 
Figure~\ref{fig:final}~[bottom~right] illustrates the actual slope of one such line.
The 2,3JA’s cost does not change as the cluster size increases, 
while the 1,3JA’s cost only gets larger. 


For each dataset, the 2,3JA algorithm's communication cost was far less than the 1,3JA algorithm's, 
a fact due entirely to the reduced output size. 
If the intermediate aggregator combined some number of KVPs ($n$) into one, 
and that KVP produced $m$ tuples in the aggregated final output, 
then $n*m$ tuples ($n$ identical sets of $m$ tuples) would be output in the unaggregated result.

\begin{figure}[h]
  \includegraphics[width=8.5cm]{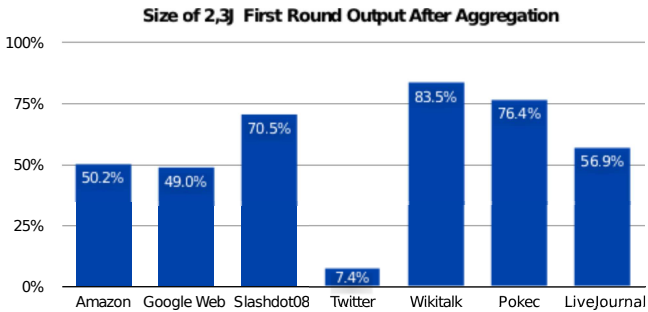} 
  \caption{This chart shows the size of the intermediate aggregation of the first round of the 2,3JA, 
	as a percentage of the size of the first two-way join.}
   \label{fig:size_firstround}
\end{figure}
\begin{figure}[h]
  \includegraphics[width=8.5cm]{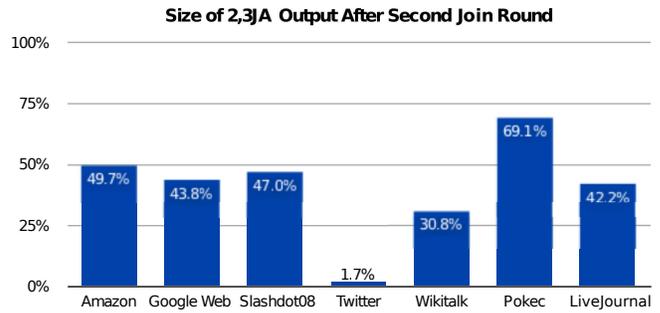} 
  \caption{This chart shows the size of the output of the 2,3JA, 
	measured as a percentage of the output size of the 1,3J, where aggregation is not used.}
  \label{fig:size_output}
\end{figure}

The benefits of this (as shown in Figure~\ref{fig:size_firstround})  were transferred over to the 2,3JA's second join round, producing a reduction in output size. The exact reduction in size is shown in Figure~\ref{fig:size_output}.  As an example, the output from the primary aggregation round on the Pokec dataset is 76.4\% of the size of first two-way join, 
a little larger than the average. 
This benefit carries over to the algorithm's output: 
the 2,3JA's Pokec output is 69.1\% the size of the 1,3J's output on the same dataset. 
In comparison, the LiveJournal dataset's intermediate aggregated result is 56.9\% the size of the first two-way join; 
the final 2,3JA output is 42.2\% the size of the 1,3J output.

\noindent

\section{Conclusions}
\label{conclusions}

We focused on three-way joins for MapReduce. 
Such joins are especially useful for friends-of-friends analysis and triangle computation 
in social networks. 
We considered the algorithm of \cite{AU10,AU11} (1,3J) versus a simple cascade of two-way joins (2,3J). 
The communication cost of 1,3J is dependent on the number of reducers; 
the more the number of reducers used in the computation, 
the bigger the communication cost becomes.
On the other hand, the communication cost of 2,3J does not change when the number of reducers changes. 
We showed that 1,3J can scale much better than what was suggested in \cite{AU10,AU11}, 
often by one or two orders of magnitude.
However, when the result of the join needs to be aggregated in some way as in the case of 
matrix multiplication, then a cascade of two-way joins (2,3JA) is preferable to 1,3JA as the aggregation 
can be pushed to the intermediate results of 2,3JA significantly reducing the communication cost incurred.

\bibliographystyle{abbrv}
\bibliography{References}{}

\newpage

\noindent

\end{document}